\documentclass[a4paper,conference]{IEEEtran}

\usepackage{cite}      

\usepackage{graphicx}  

\usepackage{subfigure} 
\usepackage{amssymb}
\usepackage{graphicx}
\usepackage{multirow}
\usepackage{subfigure}
\usepackage{amssymb}
\usepackage{amsmath}
\usepackage{float}
\usepackage{rotating}
\usepackage{array}
\usepackage{color} 
\usepackage{lineno}
\usepackage{comment}

\usepackage{fancyheadings}
\newcommand{\br}{{\mathbf r}}

\newcommand{\bx}{{\mathbf x}}
\newcommand{\by}{{\mathbf y}}

\begin{document}

\title{Continuum Model for
  Nanoscale Multiphase Flows}

\author{\IEEEauthorblockN{Alexandre M. Tartakovsky}
\IEEEauthorblockA{Pacific Northwest National Laboratory\\
Richland, WA USA\\
Alexandre.Tartakovsky@pnnl.gov}
}


\maketitle

\begin{abstract}
We propose a nonlocal model for surface tension. This model, in combination with the Landau-Lifshitz-Navier-Stokes equations, describes mesoscale features of the multiphase flow, including the static (pressure) tensor and curvature dependence of surface tension. The nonlocal model is obtained in the form of an integral of a molecular-force-like function added into the momentum conservation equation. 
We present an analytical steady-state solution for fluid pressure at the fluid-fluid interface and numerical Smoothed Particle Hydrodynamics solutions that reveal the mesoscopic features of the proposed model.  
\end{abstract}

\section{Introduction}
Since the work of Young, Laplace, and Gauss, multiphase flow at the continuum scale has been modeled almost exclusively by the Young-Laplace (YL) law, imposed as a boundary condition for the Navier-Stokes (NS) equations at the fluid-fluid interface. In this work, we propose an alternative continuum description of the multiphase flow that recovers nanoscale physics  at small (molecular) scales and reproduces the YL law at  larger scales.   

At the molecular scale, multiphase flow is traditionally described by the equations of molecular dynamics (MD). 
Experiments and MD simulations show a sharp density and pressure drop across the fluid-fluid interface in the region of about 10 nanometers. 
Outside of this region, the pressure satisfies the YL law, i.e., the pressure difference across the interface is linearly proportional  to the interface curvature.  

Another molecular-scale feature of fluid-fluid interfaces is the surface tension dependence on the curvature radius for curvature radii smaller than 10 nanometers. As the curvature radius increases, surface tension asymptotically approaches its ``macroscopic'' value $\sigma_0$. 
Therefore, one can conclude that the NS-YL model is suitable for describing interfacial dynamics on scales larger than 10 nanometers given that the thermal fluctuations are properly accounted for \cite{Landau1987}. 
Here, we propose a continuum nonlocal surface tension model that when combined with the NS equation, is capable of capturing molecular-scale features, such as pressure drop across the interface and surface tension dependence on the curvature radius. 
On the molecular scale, surface tension results from the broken symmetry in the molecular interactions near the interface, i.e., the molecular forces acting between like molecules differ from forces acting between unlike molecules. In the proposed model, the surface tension is modeled via a nonlocal term in the form of an integral of molecular-force-like function added to the momentum conservation equation.    
   
We present a semi-analytical steady-state solution for the fluid pressure across a fluid-fluid interface. This solution shows mesoscopic features for large curvatures, i.e., a non-linear curvature dependence of the pressure difference across the interface.  For small curvatures, the analytical solution demonstrates macroscale features away from the interface (i.e., static stress tensor is isotropic and satisfies the YL law) and mesoscale features close to the interface (i.e., an anisotropic static stress tensor).   Finally, we demonstrate that the model accurately describes thermally-driven fluctuations of the interface.

\section{Non-local model }

Flow of two or more incompressible fluids in the presence of thermal fluctuations can be described by the Landau-Lifshitz-Navier-Stokes (LLNS) equations \cite{Landau1987} written for each fluid phase $\alpha$ in the form of the continuity equation: 

\begin{equation}
\frac{D \rho_\alpha }{D t}=-\rho_\alpha \nabla \cdot \mathbf{v}_\alpha ,  \quad \mathbf{x} \in \Omega_\alpha
\label{Eq-Cont}
\end{equation}
and the momentum conservation equation: 
\begin{equation}
{\rho_\alpha}\frac{{{D}{{\bf{v}}_\alpha}}}{{{D}t}} =  -  \nabla P_\alpha + \nabla \cdot   \boldsymbol{\tau}_\alpha + {\rho_\alpha}{\bf{g}} + \nabla \cdot \mathbf{s}_\alpha  \quad\mathbf{x}\in\Omega_\alpha,
\label{Eq-Mom}
\end{equation}
where $\Omega_\alpha$ and $\Omega_\beta$ are the domains occupied by the $\alpha$ and $\beta$ fluids, $\Omega=\Omega_\alpha \cup\Omega_\beta$ is the total computational domain and $\Omega_\alpha \cap \Omega_\beta=0$, $ {\bf{v}}_\alpha$ is the velocity vector,  
$ \boldsymbol{\tau}_\alpha =  \left[ { \mu_\alpha ( { {{\nabla {\bf{v}}}_\alpha } } + {{ {\nabla {\bf{v}}}_\alpha }^{{\rm{T}}}}}  )  \right]  $
 is the viscous stress tensor, $\rho_\alpha$ is the density, $P_\alpha$ is the pressure, $\mu_\alpha$ is the dynamic viscosity, ${\bf{g}}$ is the gravitational acceleration, and ${\rm{D}}/{\rm{D}}t=\partial/\partial t + {\bf v}_\alpha \cdot \nabla $ denotes a total derivative. 
 
Fluctuations in velocity are caused by the random stress tensor
\begin{equation}
\mathbf{s}_l = \gamma_l \boldsymbol{\xi},
\end{equation}
where $\boldsymbol{\xi}$ is a random symmetric tensor and $\gamma_l$ is the strength of the noise.
The random stress is related to the viscous stress by the fluctuation-dissipation theorem \cite{Landau1987}. For incompressible and low-compressible fluids, the covariance of the stress components is:
\begin{equation}
\overline{{s^{in}_l(\mathbf{x}_1,t_1)s^{jm}(\mathbf{x}_2,t_2)}}=
\gamma^2_l \delta(\mathbf{x}_1-\mathbf{x}_2)\delta(t_1-t_2),
\label{eq:randomstress_squared}
\end{equation}
where $\gamma^2_l = 2 \mu_l k_B T \delta^{ij}\delta^{nm}$, $k_B$ is the Boltzmann constant, $T$ denotes the temperature, $\delta(z)$ is the Dirac delta function, and $\delta^{ij}$ is the Kronecker delta function.

The YL boundary condition for pressure and velocity is imposed at the fluid-fluid-interface
\begin{equation}\label{Young-Laplace}
(P_\alpha-P_\beta)\mathbf{n}= (\boldsymbol{\tau}_\alpha-\boldsymbol{\tau}_\beta)\cdot\mathbf{n} +\kappa \sigma_0 \mathbf{n}.
\end{equation}

Many grid-based methods for the the multiphase NS equations rely on the Continuum Surface Force model  \cite{brackbill1992}, which replaces Eqs. (\ref{Eq-Mom}) and (\ref{Young-Laplace}) with: 

\begin{equation}
{\rho}\frac{{{D}{{\bf{v}}}}}{{{D}t}} =  -  \nabla P + \nabla \cdot   \boldsymbol{\tau} + \rho{\bf{g}} + \nabla \cdot \mathbf{s} +  \mathbf{F}
\quad\mathbf{x}\in\Omega.
\label{Eq-Mom-Brackbill}
\end{equation}

Assuming that the fluid is inviscid, the surface force $\mathbf{F}$ can be expressed in terms of the surface tension as: 

\begin{equation}\label{SVF}
\mathbf{F} = \sigma_0 \kappa \nabla \phi,
\end{equation}
where $\phi$ is the color function  

\begin{equation}
\phi(\mathbf{x}) =  \left \{ \begin{array}{rcl}
     1, & \mathbf{x}\in\Omega_\beta,\\
     0, & \mathbf{x}\in\Omega_\alpha.
   \end{array} \right.
\label{equ:color_func}
\end{equation}

In this paper, we propose a new model for surface tension, where $\mathbf{F}$ is replaced with
\begin{equation}
  \mathbf{F}(\mathbf{x}) =  -   \int_\Omega  s(\bx,\by) f_\varepsilon(|\bx - \by|) \frac{\bx-\by}{|\bx-\by|} d \by \quad\mathbf{x}\in\Omega,
  \label{nonlocal-force}
\end{equation} 
where  $s(\bx,\by)$ is the force strength and $f_\varepsilon(|\bx - \by|)$  is the force shape function that is negative for small $|\bx - \by|$, positive for large $|\bx - \by|$, and is zero for $|\bx - \by|\ge \varepsilon$.

The force strength is defined as
\begin{equation}
s(\bx,\by)
 = \left\{ {\begin{array}{*{20}{c}}
   s_{\alpha\beta}, &  \; \mathbf{x} \in \Omega_\alpha \quad \text{and} \quad  \mathbf{y} \in \Omega_\beta,  \\
   s_{\alpha\alpha}, &  \; \mathbf{x} \in \Omega_\alpha \quad \text{and} \quad  \mathbf{y} \in \Omega_\alpha,  \\
   s_{\beta\beta}, &  \; \mathbf{x} \in \Omega_\beta \quad \text{and} \quad  \mathbf{y} \in \Omega_\beta.  \\
\end{array}} \right.
\label{Eq-SPH_Int}
\end{equation}

An essential feature of the proposed model is that it recovers the stress tensor at the fluid-fluid interface. In the proposed model, the static stress tensor is 

\begin{equation}
\mathbf{T} = 
\left(\begin{array}{ccc}
P+T^F_N & 0 & 0
\\
0 & P+T^F_T & 0
\\
0 & 0 & P+T^F_T
\end{array} \right),
\end{equation}
where $T^F_N$ and $T^F_N$ are the normal and tangent stress components generated by $\mathbf{F}$ in (\ref{nonlocal-force}), and $\nabla \cdot \mathbf{T} = 0$. To demonstrate the former, consider a two-dimensional infinite space ($x_1,x_2$) with flat interface parallel to $x_2$ and located at $x_1=0$, i.e., $\Omega_\alpha=(-\infty,0]\cup(-\infty,\infty)$, and $\Omega_\beta=[0,\infty)\cup(-\infty,\infty)$.  
 
Then, $P$, $T^F_N$, and $T^F_T$ can only be constants or functions of $x_1$. Under static conditions and in the absence of thermal fluctuations, Eqs. (\ref{Eq-Mom-Brackbill}) and(\ref{nonlocal-force}) yield
 
 \begin{eqnarray}
&& \frac{\partial P(x_1)}{\partial x_1} = - \frac{\partial T_N^F (x_1)}{\partial x_1} = \\ \nonumber
&& -    \int_\Omega  s(\bx,\by)   f_\varepsilon(|\bx - \by|) \frac{x_1- y_1}{|\bx-\by|} d \by = -g(x_1)
   \label{DPX}  
 \end{eqnarray}
and
 \begin{eqnarray}    
 &&\frac{\partial P(x_1)}{\partial x_2} =
- \frac{\partial T_T^F (x_1)}{\partial x_2}  =    \\ \nonumber
&&-   \int_\Omega  s(\bx,\by)  f_\varepsilon(|\bx - \by|) \frac{x_2- y_2}{|\bx-\by|} d \by = 0,
\label{DPY}
 \end{eqnarray}
where the form of $g(x_1)$ depends on the choice of  $f_\varepsilon(|\bx - \by|)$. Then, 
$T_N^F(x_1) =  T_N^F (x_1=-\infty) + \int_{-\infty}^{x_1}  g(s) ds$,
$T_{T}^F(x_1) =   T_{T}^F(x_1 = -\infty) $, and
 $P(x_1) =  P(-\infty)  -  T_{N}^F(x_1)$.
Because the stress tensor should be isotropic when away from the interface, $T_N^F (x_1=-\infty) = T_T^F (x_1=-\infty)$.  
Therefore, the total static stress $\mathbf{T}$ is anisotropic with the diagonal components:
\begin{equation}\label{TN}
T_N =  P(x_1 =-\infty) 
\end{equation}
and
\begin{equation}\label{TT}
 T_T(x_1) =  P(-\infty)  -  \int_{-\infty}^{x_1}  g(s) ds.
\end{equation}
 
 The relationship between macroscopic surface tension $\sigma_0$ and stress \cite{rowlinson2002molecular},
 \begin{equation}
 \sigma_0 = \int \limits_{-\infty}^\infty [T_N(x_1)-T_T(x_1)] dx_1 =  \int \limits_{-\infty}^\infty \int \limits_{-\infty}^{x_1}  g(s) ds dx_1,
 \end{equation}
can be integrated exactly for a flat interface, resulting in the exact relationship between $\mathbf{F}$ and $\sigma_0$:  
 
 \begin{equation}
 \sigma_0 = \frac13 (s_{\alpha \alpha} + s_{\beta \beta} -2 s_{\alpha \beta})  \int \limits_0^\infty z^3  f_\varepsilon (z)dz.
 \end{equation}

In three dimensions, similar arguments lead to 
\begin{equation}\label{T-F-Int}
\sigma_0 = \frac18 \pi (s_{\alpha \alpha} + s_{\beta \beta} -2 s_{\alpha \beta})  \int \limits_0^\infty z^4 f_\varepsilon (z)dz.
\end{equation}

Similar expression, but in terms of the intermolecular forces, can be obtained by coarse-graining the MD equations:
\begin{equation}\label{ST-MD}
\sigma_0 = \frac18 \pi  \int \limits_0^\infty z^4 g(z) [ f_{\alpha \alpha}^{LJ} (z) + f_{\beta \beta}^{LJ}  (z) - 2f_{\alpha \beta}^{LJ}  (z) ]dz,
\end{equation}
where  $g(z)$ is the radial distribution function and  $f_{\alpha \alpha}^{LJ}$, $f_{\beta \beta}^{LJ}$, and $f_{\alpha \beta}^{LJ}$ are the Lennard-Jones (LJ) forces acting between $\alpha - \alpha$, $\beta - \beta$, and $\alpha - \beta$ molecules, respectively.  Note, that unlike Eq. (\ref{T-F-Int}), Eq. (\ref{ST-MD}) is an approximation, which can be further simplified by setting $g(z)=1$. Comparing Eqs. (\ref{T-F-Int}) and (\ref{ST-MD}) reveals that $ f_\varepsilon (z)$ is a coarsened form of the molecular forces. In general, one can use LJ forces instead of $ f_\varepsilon (z)$ in Eq. (\ref{nonlocal-force}), but this would lead to very stiff equations. As the behavior of averages (density and momentum) is of primary interest here, there is no need to use LJ forces. Moreover, for computational efficiency, it is beneficial to use a differentiable $ f_\varepsilon (z)$.

Because $\sigma_0$ is positive, the force strength coefficients should satisfy $s_{\alpha \alpha } + s_{\beta \beta} > s_{\alpha \beta}$. It is convenient to assume  $s_{\alpha \alpha } = s_{\beta \beta} = 10^k s_{\alpha \beta}$  ($k>1$). Then, for a given form of  $ f_\varepsilon(|\bx - \by|)$, the $s$ coefficients can be found as a function of $\sigma_0$:

\begin{equation}\label{s11}
 s_{\alpha \alpha }   = {s}_{ \beta \beta} =\frac1{2(1-10^{-k})}  \frac{\sigma_0}{ \lambda}.
\end{equation}

 A similar analysis under static conditions can be performed for a spherical interface separating two fluids. Integration in spherical coordinates also produces an anisotropic static stress tensor at the interface and an isotropic stress tensor away from the interface. For the radius of curvature larger than $\varepsilon$, the components  of the isotropic stress tensor satisfy the YL equation. 
Note that for the considered problems (two fluids under static conditions separated by a flat or spherical interface), Eq. (\ref{Eq-Mom-Brackbill}) will predict an isotropic stress tensor.
Another fundamental difference between the nonlocal model and the YL law is that the former has an internal length scale $\varepsilon$, while the latter does not have any internal length scale.  Because of this, the YL law predicts the same behavior (for same dimensionless numbers) regardless of the problem's length scale. The described above analytical solutions and numerical results in the following section  show that the nonlocal model behavior depends on the ratio of its internal length scale  to the characteristic length of the problem.

\section{Smoothed Particle Hydrodynamics solution of the nonlocal LLNS equation}

In  \cite{Lei2016PRE}, we derived a discretized form of Eqs. (\ref{Eq-Cont}), (\ref{Eq-Mom-Brackbill}), and (\ref{nonlocal-force}) by adding pairwise interaction forces 

\begin{equation}
\mathbf{F}^{int}_i = \frac{1}{V_i}\sum_{j=1}^N {\bf{F}}_{ij}^{int}\hspace{0.5cm}\text{and}
\hspace{0.5cm} {\bf{F}}_{ij}^{int} = -s(\br_i,\br_j)f_\varepsilon(r_{ij}) \frac{\mathbf{r}_{ij}}{r_{ij}},
\label{int-force}
\end{equation}  
to the Smoothed Particle Hydrodynamics (SPH) form of the LLNL equations. In that work, the domain was discretized with points (or particles) with positions $\mathbf{r}_i$, $\mathbf{F}_i$ is the ``surface tension'' force acting on point $i$, $\mathbf{f}_{\varepsilon}(r_{ij})$ is the molecular-like force acting between points $i$ and $j$, ${\bf r}_{ij} = {\bf r}_i - {\bf r}_j $,  and $r_{ij} = |{\bf r}_{ij}| $. Various forms of $\mathbf{f}_{\varepsilon}(r_{ij})$ were suggested in \cite{Tart-PFSPH}. In \cite{Lei2016PRE}, we used 
\begin{equation}
\mathbf{f}_{\varepsilon}(r_{ij})= r_{ij} \left[-Ae^{-\frac{r_{ij}^2}{2\varepsilon_0^2}}
+ e^{-\frac{r_{ij}^2}{2 \varepsilon^2}} \right],
\label{eq:pair_interaction}
\end{equation}
where $\varepsilon_0 = \varepsilon/2 = dp/2<h$ ($h$ is the support of the SPH weighting function, and $dp$ is the average spacing between SPH particles or the lattice size if the particles are initially placed on a uniform lattice) was found to maintain a relatively uniform particle distribution during simulations. Also, this choice of parameters makes the range of  $\mathbf{f}_{\varepsilon}(r_{ij})$ to be approximately equal to $\varepsilon$, which is of the same order as $h$.

\begin{figure}[!h]
\includegraphics*[scale=0.3]{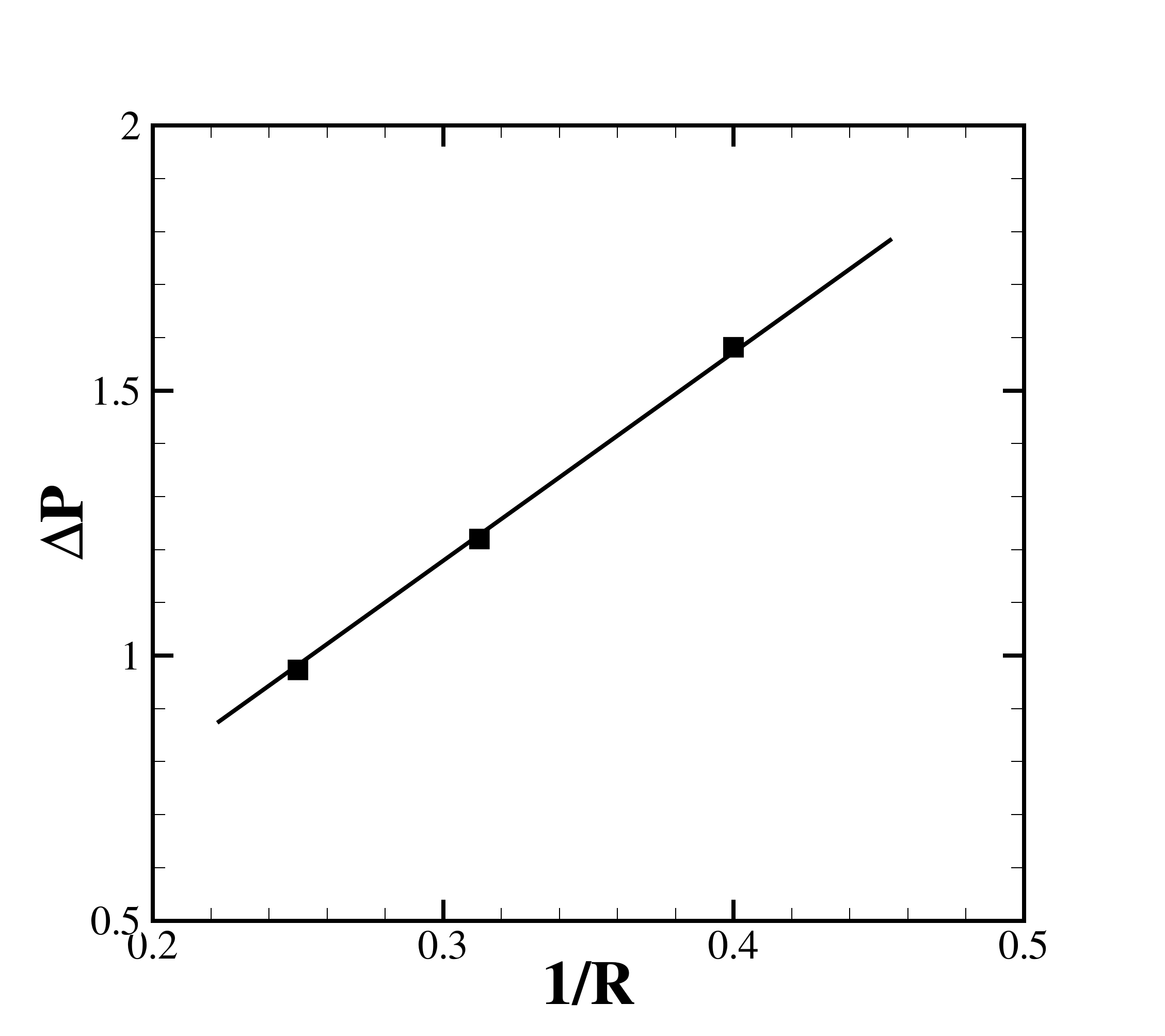}
\caption{ Pressure difference between the inside and outside domain of the droplet
  of radius $R = 2.5$, $3.2$, and $4.0$ with $n_{eq} = 64.0$ and $k_BT = 0.001$.
  The solid line represents the analytical prediction from the 
    YL relationship with $\sigma_0 = 2.0$. From  \cite{Lei2016PRE}.
}
\label{fig:droplet_surface_tension_a}
\end{figure}

In \cite{Lei2016PRE}, we also computed the surface tension of three-dimensional droplets of different sizes.  Figure \ref{fig:droplet_surface_tension_a} shows that the pressure follows the YL law for droplets with radii at least 2.5 times larger than $h$.  

\begin{figure}[!h]
\includegraphics*[scale=0.3]{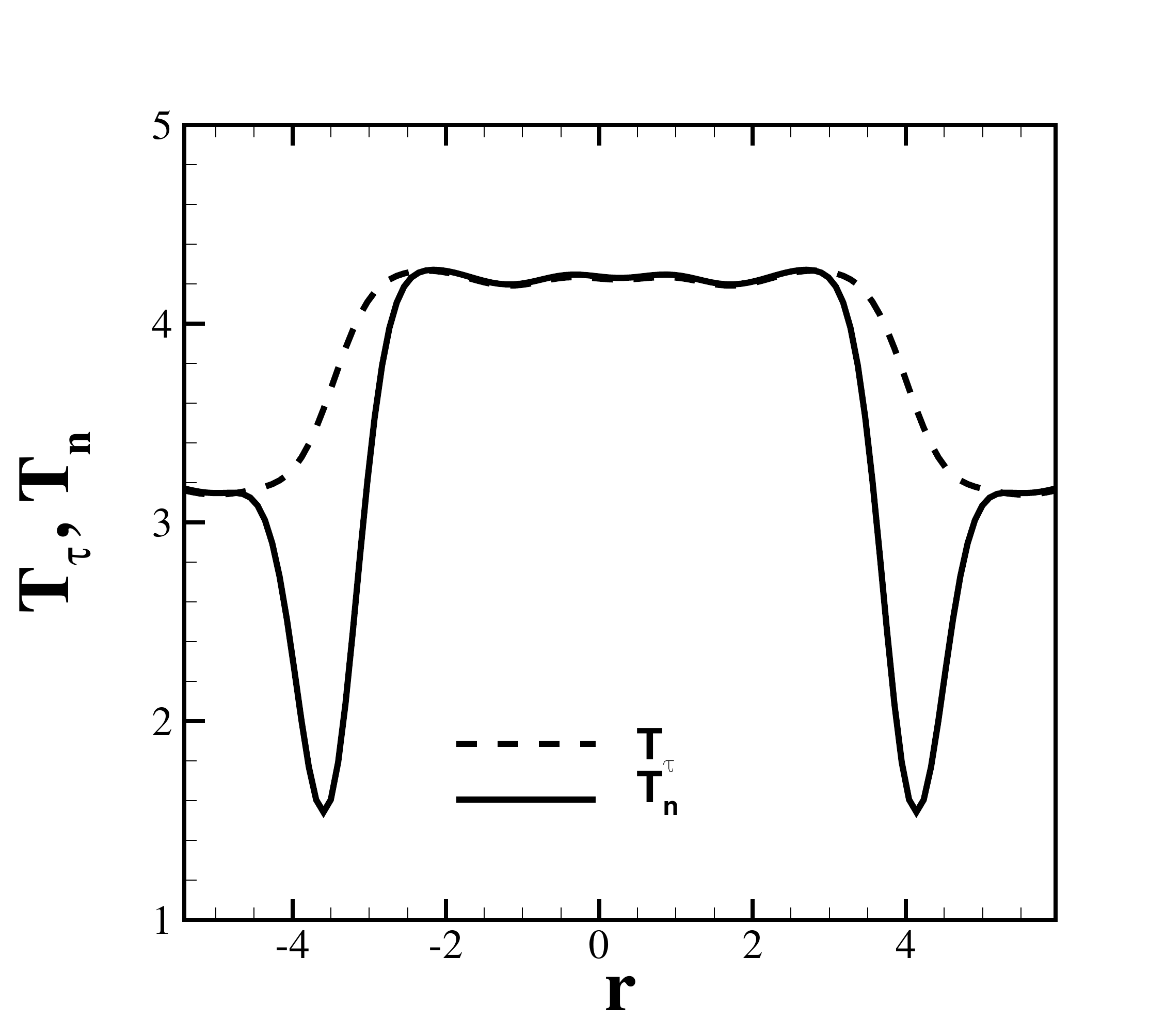}
\caption{
 Normal and tangential hardy stress along the radial direction of the droplet
of radius $R = 4.0$. From \cite{Lei2016PRE}.
}
\label{fig:droplet_surface_tension_b}
\end{figure}

Figure~\ref{fig:droplet_surface_tension_b} displays the static stress components $T_{n}$
and $T_{\tau}$ for a droplet with $R=4h$. Note that the YL assumes that the static stress is isotropic.  

\begin{figure}[!h]
\includegraphics*[scale=0.3]{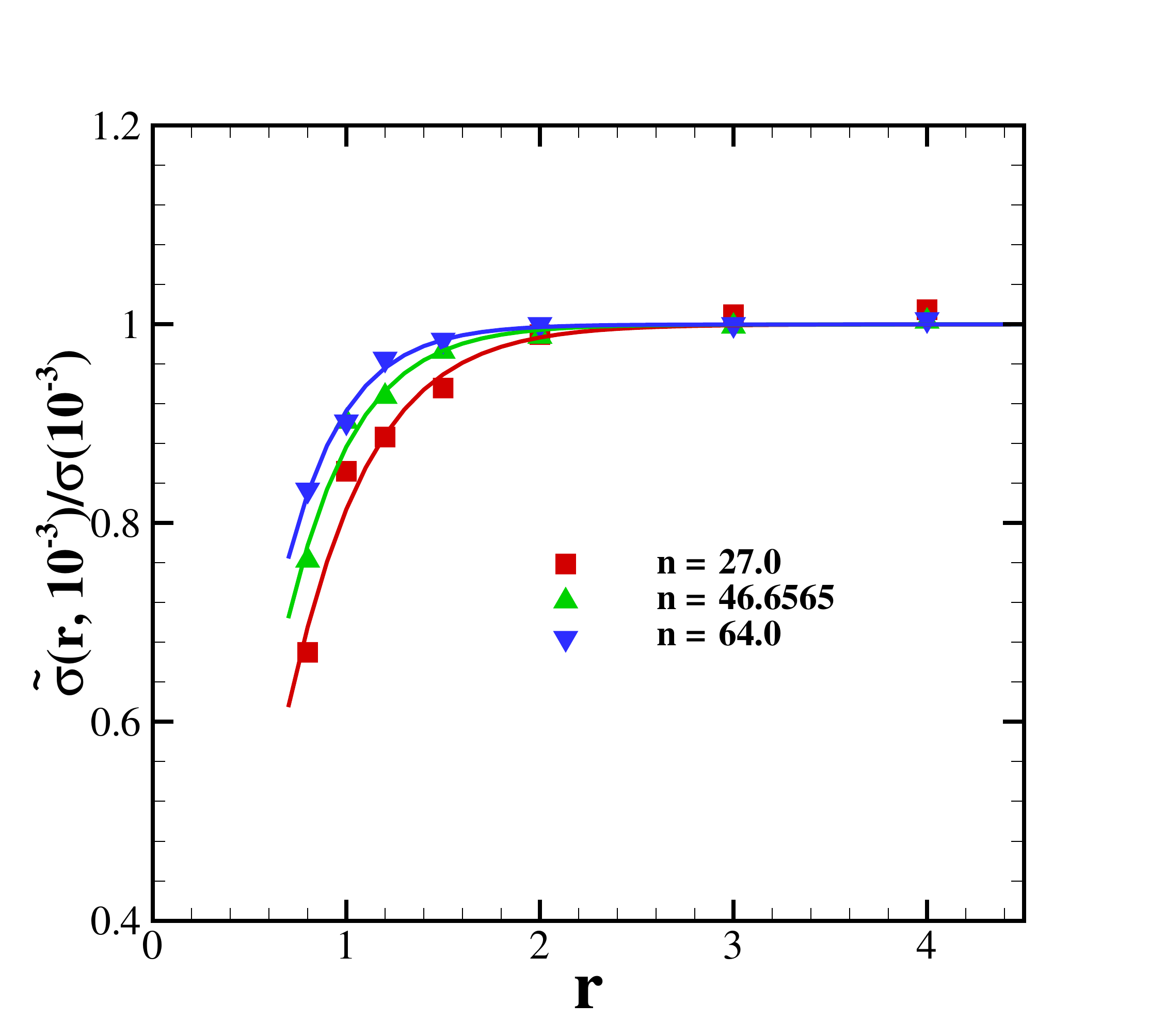}
\caption{
 Normalized surface tension $\sigma(R)/\sigma_0$ computed from droplets
of different radii $R$ with $k_BT = 0.001$ and $\sigma_0 = 2.0$. From  \cite{Lei2016PRE}.
}
\label{fig:droplet_surface_tension_c}
\end{figure}

Figure \ref{fig:droplet_surface_tension_c} depicts the surface tension computed from droplet simulations as 
 \begin{equation}
 \sigma(R) = \int \limits_{0}^\infty [T_N(r)-T_T(r)] dr,
 \end{equation}
where $R$ the droplet radius and the integration is performed along the line perpendicular to the droplet surface. The simulations were performed with different average particle number density $n_{eq}$ (the number of particles per unit volume) and $R$ ranging from $0.8h$ to $4h$. For all considered $n_{eq}$ and $R<2h$,
$\sigma(R)$ decreases with decreasing $R$ from its macroscopic value $\sigma_0$. Such behavior has been observed in nanoscale droplets with the radius smaller than several nanometers depending on the fluid type \cite{kashchiev2003determining}. Therefore, to quantitatively describe the nanoscale behavior, $\varepsilon$ (and $h$) should be on the order of  several nanometers.

\begin{figure}[!h]
\includegraphics*[scale=0.3]{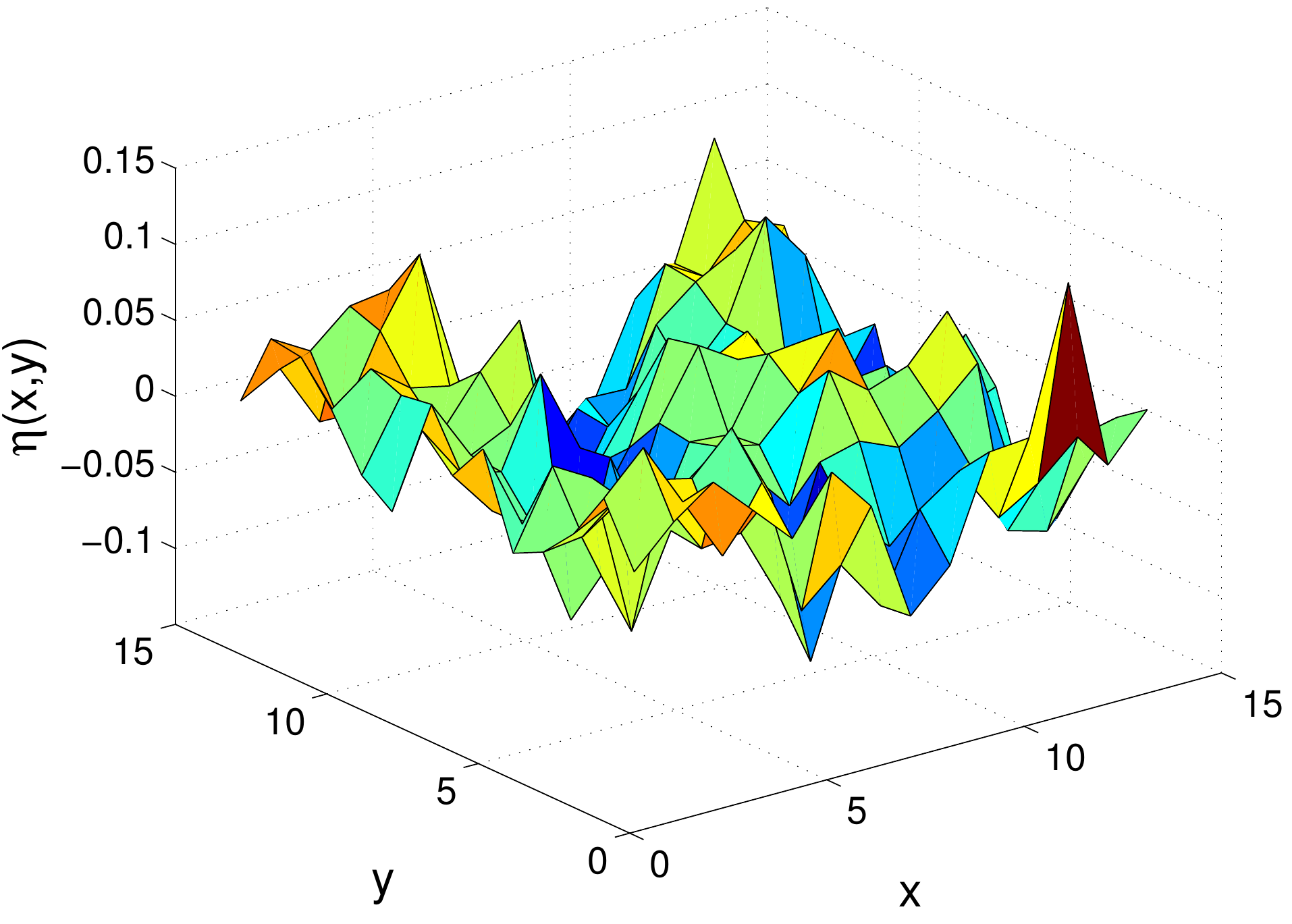}
\caption{
Instantaneous fluid height near the interface with $g=0$ at $k_BT = 0.01$. From  \cite{Lei2016PRE}.
}
\label{fig:CWT_neutral}
\end{figure}
In \cite{Lei2016PRE}, we also modeled the interfacial capillary waves on the fluid-fluid interface $\eta(x,y)$ caused by thermally induced velocity fluctuations. 
According to the capillary wave theory (CWT)  \cite{Buff_Lovett_PRL_1965, Evans_ADP_1979},  in the absence of gravity,  the Fourier modes  $\hat{\eta}(\mathbf{q})$ of $\eta(x,y)$ are given by
\begin{equation}
\left<\hat{\eta}(\mathbf{q})^2\right> = \frac{k_BT}{\sigma
  \left\vert \mathbf{q}\right\vert^2 L^2},
\label{eq:CWT}
\end{equation}
where $L\times L$ is the lateral interface domain.
When gravity $g$ acts in the $z$ direction, the variance of $\hat{\eta}(\mathbf{q})$ is
\begin{equation}
\left<\hat{\eta}(\mathbf{q})^2\right> = \frac{k_BT}{\sigma \left\vert\mathbf{q}\right\vert^2 L^2 + (\rho_{\alpha}-\rho_{\beta}) g L^2}.
\label{eq:CWT_gravity}
\end{equation}

\begin{figure}[!h]
\includegraphics*[scale=0.3]{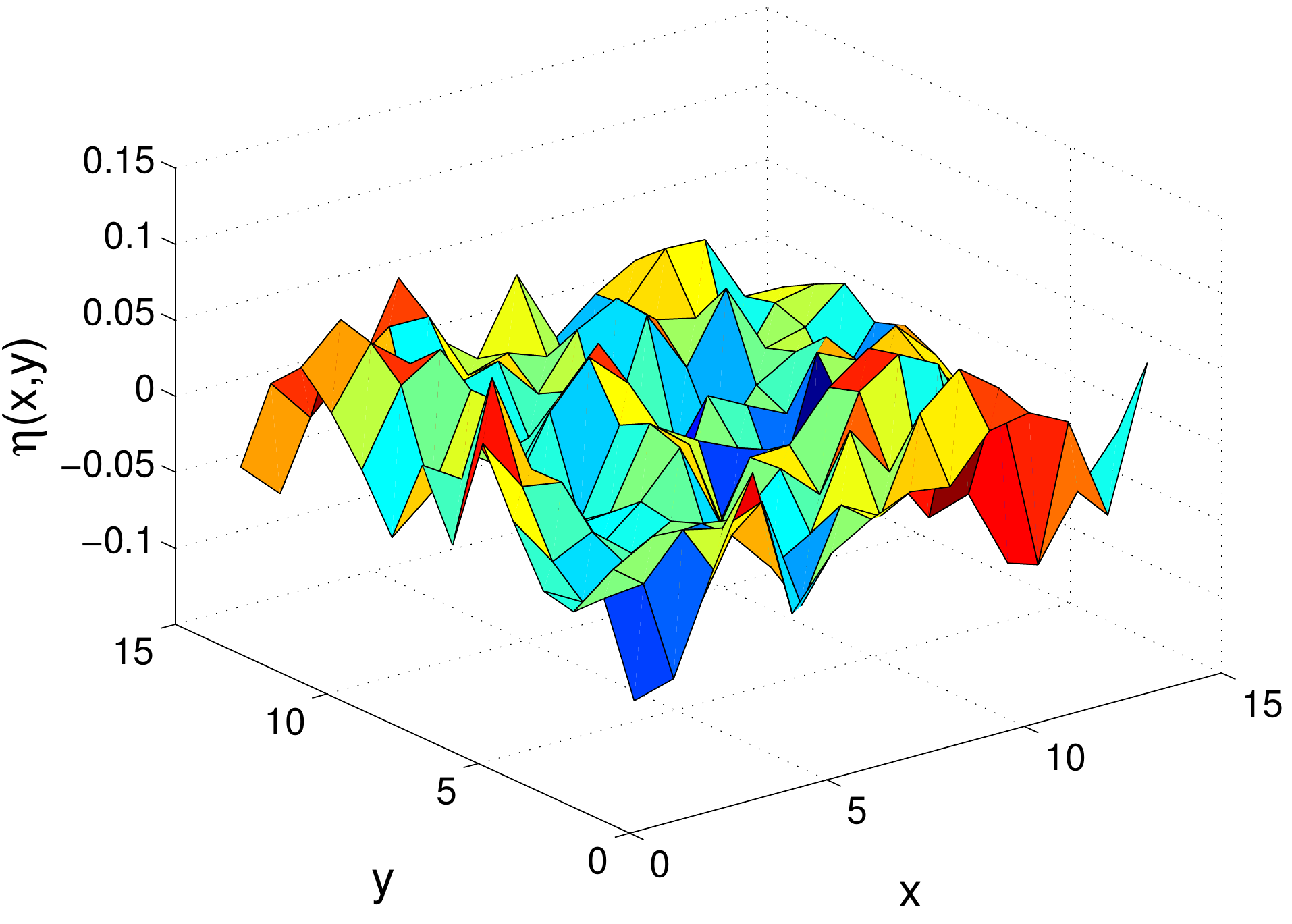}
\caption{Instantaneous fluid height near the interface with
  $\rho_{\alpha} = 32.0$,
  $\rho_{\beta} = 64.0$, and $g = 0.04$ at $k_BT = 0.01$. From  \cite{Lei2016PRE}.
}
\label{fig:CWT_gravity_a}
\end{figure}
We considered three systems, including (I) a system without gravity; (II) a system with a heavier fluid overlaying a lighter fluid with 
$\rho_{\alpha} = 64.0$, $\rho_{\beta} = 32.0$, and gravity $g = 0.04$; and (III) a system with a lighter fluid on top of a heavier fluid wth  
$\rho_{\alpha} = 32.0$, $\rho_{\beta} = 64.0$, and $g = 7.5\times10^{-3}$.

\begin{figure}[!h]
\includegraphics*[scale=0.3]{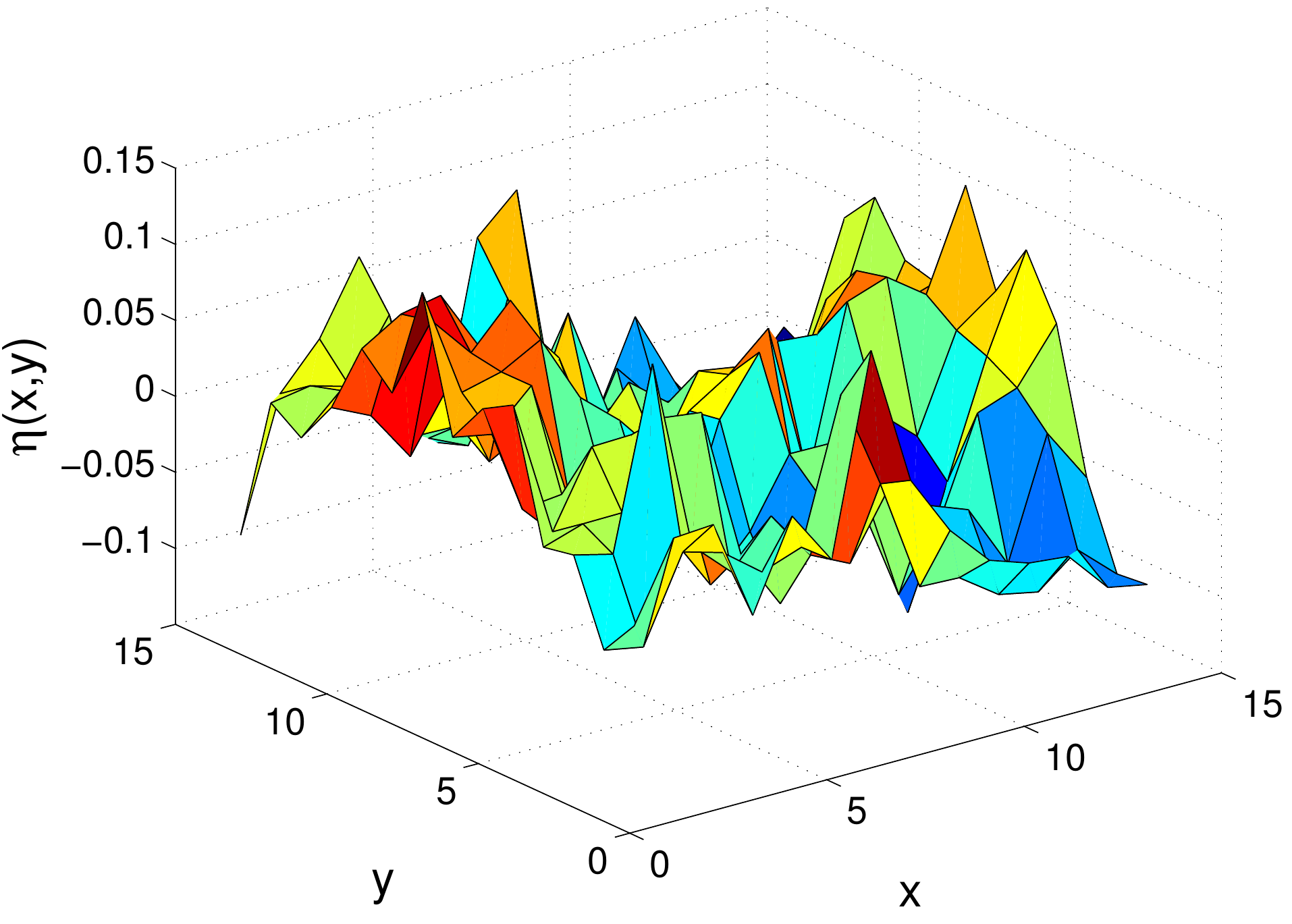}
\caption{
 Instantaneous fluid height near the interface with
$\rho_{\alpha} = 64.0$, $\rho_{\beta} = 32.0$, and $g = 7.5\times10^{-3}$ at $k_BT = 0.01$. From  \cite{Lei2016PRE}.
}
\label{fig:CWT_gravity_b}
\end{figure}

Figures \ref{fig:CWT_neutral}-\ref{fig:CWT_gravity_b} show the instantaneous
interface $\eta(x,y)$ for Cases I-III, respectively. 
As expected, gravity increases interface fluctuations in Case II and decreases fluctuations in Case III.  

\begin{figure}[!h]
\includegraphics*[scale=0.3]{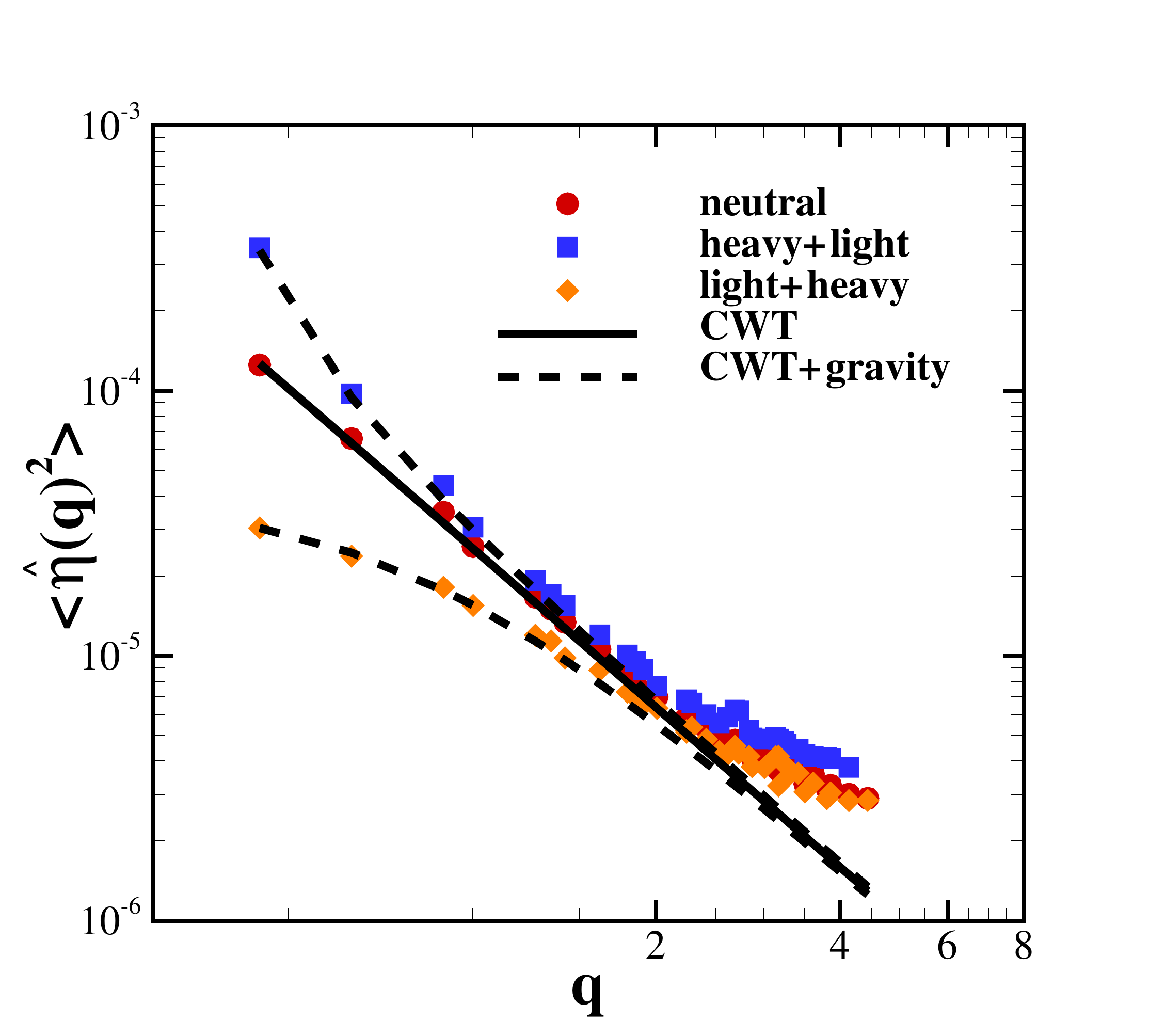}
\caption{
The capillary wave spectra measured at different gravity. From  \cite{Lei2016PRE}.
}
\label{fig:CWT_gravity_c}
\end{figure}
\begin{figure}[!h]
\includegraphics*[trim = 5mm 0mm 5mm 20mm,clip,scale=0.3]{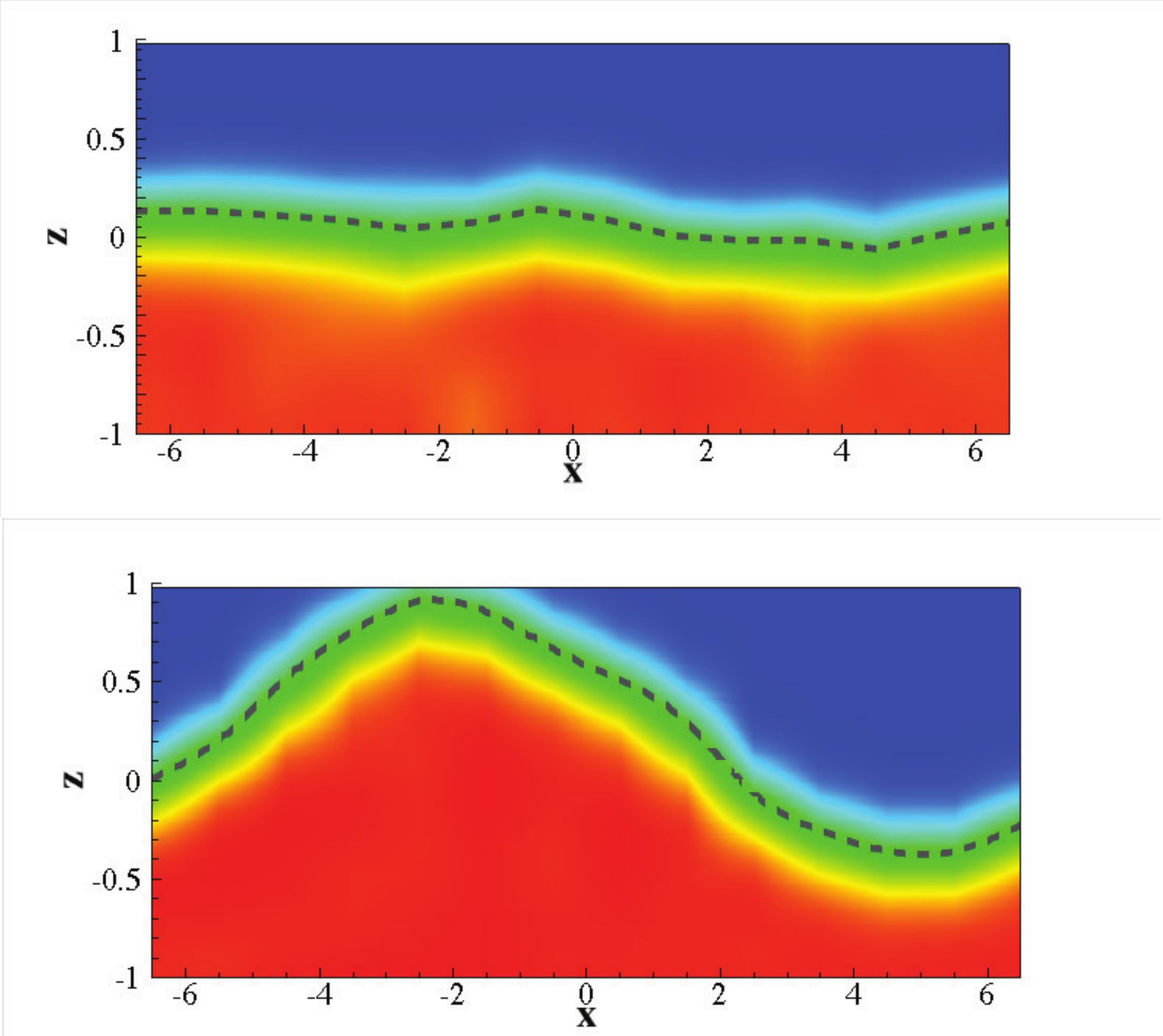}
\caption{
Instantaneous density field near the fluctuation interface
for $g = 0.0075$ (upper) and $g = 0.014$ (lower). For $g = 0.0075$ satisfying Eq. (\ref{eq:Ray_instability}), the stable interface is maintained. 
For $g = 0.014$ violating Eq. (\ref{eq:Ray_instability}), Rayleigh instability develops. From  \cite{Lei2016PRE}.}
\label{fig:CWT_gravity_d}
\end{figure}

Figure~\ref{fig:CWT_gravity_c} shows that the variance of $\hat{\eta}(\mathbf{q})$
for Cases I-III found from SPH simulations is in a good agreement 
with the predictions from Eqs. (\ref{eq:CWT}) and (\ref{eq:CWT_gravity}) for 
$|\mathbf{q}| \le \frac{2\pi}{5h}$. 
In Case III, to prevent the Rayleigh
instability from developing, we chose $g$ and $\sigma_0$ to satisfy
\begin{equation}
\sigma_0 q_0^2 > (\rho_{\beta} - \rho_{\alpha})g,
\label{eq:Ray_instability}
\end{equation}
where $q_0 = \frac{2\pi}{L}$ is the smallest stable wave number.  Figure~\ref{fig:CWT_gravity_d} demonstrates that Rayleigh instability develops when  Eq.~(\ref{eq:Ray_instability})
is violated.

\section{Conclusion}
In this paper,  we proposed a multisclale nonlocal momentum conservation equation for multiphase flow with internal scale $\varepsilon$. 
We demonstrated that this model captures important nanoscale features of interfaces, including the static stress tensor at the interface and temperature driven  interface fluctuations. Our results show that the proposed model predicts the curvature-dependent surface tension of interfaces with the radius of curvature of the same order or smaller than $\varepsilon$. For interfaces with a larger radius of curvature, the model recovers macroscopic behavior with (macroscopic) surface tension and pressure difference across the interface, satisfying the YL law.


\section*{Acknowledgment}
This work was supported by the U.S. Department of Energy (DOE) Office of Science, Office of Advanced Scientific Computing Research as part of the New
Dimension Reduction Methods and Scalable Algorithms for Nonlinear Phenomena project. Pacific Northwest National Laboratory is operated by Battelle for the DOE under Contract DE-AC05-76RL01830.



\bibliographystyle{IEEEtran.bst}

%
%
%

\end{document}